# Plate motion in sheared granular fault system

Ke Gao[1,2*], Robert A. Guyer[2,3], Esteban Rougier[2], and Paul A. Johnson[2]

[1]Department of Earth and Space Sciences, Southern University of Science and Technology, Shenzhen, China

[2]Geophysics, Los Alamos National Laboratory, Los Alamos, NM, USA

[3]Department of Physics, University of Nevada, Reno, NV, USA

## Abstract

Plate motion near the fault gouge layer, and the elastic interplay between the gouge layer and the plate under stick-slip conditions, is key to understanding the dynamics of sheared granular fault systems. Here, a two-dimensional implementation of the combined finite-discrete element method (FDEM), which merges the finite element method (FEM) and the discrete element method (DEM), is used to explicitly to simulate a sheared granular fault system. We focus on investigating the influence of normal load, driving shear velocity and plate stiffness on the velocities and displacements measured at locations on the upper and lower plates just adjacent to the gouge in the direction parallel to the shear direction (*x* direction). The simulations show that at slips the plate velocities are proportional to the normal load and may be inversely proportional to the square root of the plate's Young's modulus; whereas the driving shear velocity does not show distinct influence on the plate velocities. During stick phases, the velocities of the upper and lower plates are respectively slightly greater and slightly smaller than the half of the driving shear velocity, and are both in the same direction of shear. The shear strain rate of the gouge is calculated from this velocity difference between the upper and lower plate during stick phases and thus the gouge effective shear modulus can be calculated. The results show that the gouge effective shear modulus increases proportionally with normal load, while the influence of shear velocity and plate stiffness on gouge effective shear modulus is minor. The simulations address the dynamics of a laboratory scale fault gouge system and may help reveal the complexities of earthquake frictional dynamics.

## Keywords

# 1. Introduction

The devastating potential of earthquakes to society calls for a thorough understanding of earthquake frictional dynamics. The fault gouge, an ensemble of solid granular particles created by comminution – the fragmentation and wearing of fault blocks – plays a key role in the macroscopic stability of a fault (Dorostkar et al., 2017b; Marone et al., 1990). Therefore, stick-slip cycles (or simply stick-slip) in sheared granular fault systems have been intensively studied in recent years both in laboratory experiments and numerical simulations (e.g., Gao et al., 2018; Geller et al., 2015; Johnson and Jia, 2005; Johnson et al., 2008; Marone, 1998; Marone et al., 1990; Passelègue et al., 2016; Scuderi et al., 2017). In these studies, the model employed generally consists of granular gouge compressed and sheared by confining plates which play the role of fault blocks. However, significant attention has been paid to gouge kinematics and mechanics, with few details being reported regarding the motion of the plates, especially in places adjacent to the gouge. This near gouge plate motion not only directly controls the output of acoustic signals that may serve as harbingers for earthquake prediction, but also reflects the physical and dynamic properties of the fault system such as seismic moment and frequency content of the radiated waves (Gao et al., 2019; Jackson and McKenzie, 1988; Johnson et al., 2013; Rivière et al., 2018; Rouet-Leduc et al., 2018; Rouet-Leduc et al., 2017; Siman-Tov and Brodsky, 2018; Taylor and Brodsky, 2019). Therefore, investigating stick-slip-induced near gouge plate motion in sheared granular fault systems is key to improve the understanding of the complex mechanisms of fault friction and may also shed light on ground motion and earthquake hazard.

Numerical simulation is widely used to simulate single fault or fault patches simply because of its ease of implementation and capability of analyzing granular fault at a level of spatial and temporal resolution not accessible experimentally (de Arcangelis et al., 2011). Among the many numerical methods available, the discrete element method (DEM) has been the most widely applied (e.g., Dorostkar et al., 2017a; Dorostkar et al., 2017b; Ferdowsi et al., 2013; Griffa et al., 2011; Mair and Hazzard, 2007; Wang et al., 2017). In classic DEM models, the granular fault gouge is commonly represented by a pack of rigid particles, and the representation of the confining plates is simplified by a set of bonded particles (Abe and Mair, 2005; Dorostkar et al., 2017b; Ferdowsi et al., 2014; Griffa et al., 2013; Mair and Abe, 2008) (e.g., Figure S1a of the Supplementary Material). As a result, by using DEM it is challenging to capture detailed deformation and motion within both the particles and plates (Dratt and Katterfeld, 2017; Ma et al., 2016). Particularly in DEM models, since each plate behaves as a rigid body, the spatial variation of plate motion along the gouge in response to stick-slip is difficult to acquire.

From a computational mechanics viewpoint, a granular fault system is essentially a combination of continua (each individual plate and particle) and discontinua (particle-particle and particle-plate interactions). Considering this, a numerical tool such as the combined finite-discrete element method (FDEM) (Munjiza, 1992, 2004; Munjiza et al., 2011; Munjiza et al., 2014), which merges finite element based analysis of continua with discrete element based analysis of discontinua, provides a natural solution for such problem. In a FDEM realization of the granular fault system (Figure S1b), each plate and each particle is represented by a discrete element, which allows for tracking of its motion and interactions with neighboring objects. Furthermore, each discrete element is discretized into finite elements which allows for describing its deformation in response to external forces. Therefore, by utilizing FDEM one can obtain an explicit representation of a granular fault system and thus be capable of obtaining detailed information regarding internal gouge behavior and gouge-plate motion during the full stick-slip cycle.

In this paper, based on the FDEM simulations of a granular fault system, we explore the motion of the gouge and plates during the stick-slip cycles. This analysis is conducted for models featuring different plate stiffness and subjected to different normal loads and shear velocities. We first provide a brief introduction to the numerical method including the theory of FDEM and model setup. Then the motion of both the upper and lower plates in stick and slip phases as a function of velocity and displacement in directions parallel to the shear direction (*x* direction) are analyzed. The influence of changes of the normal load, shear velocity and plate stiffness on the plate motion is demonstrated and the gouge effective shear modulus interpreted from the plate motion is presented. We discuss a possible scaling relationship for the plate motion and gouge shear modulus with respect to plate stiffness, normal load and shear velocity and conclude.

## 2. Numerical methods

### 2.1. FDEM in a nutshell

The FDEM was originally developed by Munjiza in early 1990s to simulate the material transition from continuum to discontinuum (Munjiza, 1992). The essence of this method is to merge the algorithmic advantages of DEM with those of the finite element method (FEM). The main theory of FDEM involves the algorithms of governing equations, deformation description, contact detection, and contact interaction (Lei et al., 2016; Munjiza et al., 2006).

The general governing equation of the FDEM is (Munjiza, 2004)

$$\mathbf{M}\ddot{\mathbf{x}} + \mathbf{C}\dot{\mathbf{x}} = \mathbf{f} , \qquad (1)$$

where **M** is the lumped mass matrix, **C** is the damping matrix, **x** is the displacement vector, and **f** is the equivalent force vector acting on each node. An explicit time integration scheme based on a central difference method is employed to solve Eq. (1) with respect to time to obtain the transient evolution of the system. Deformation of finite elements is described by a multiplicative decomposition based formulation, which allows for a detailed capture of material deformation (Munjiza et al., 2014). The contact detection between discrete elements is conducted using the MRCK (Munjiza-Rougier-Carney-Knight) algorithm (Rougier and Munjiza, 2010), which determines whether any two given elements, one called the contactor and the other one the target, share at least one search cell. After processing the contact detection, a list that contains all the pairs of elements potentially in contact is established and sent for contact interaction analysis. A penalty function based contact interaction algorithm is used to calculate the contact forces between contacting elements (Munjiza, 2004; Munjiza et al., 2011). Detailed calculation of normal and tangential contact forces is demonstrated in Text S1 of the Supplementary Material.

It is beyond the scope of the present paper to provide a complete description of the above principles; however, details of these can be found in FDEM monographs (Munjiza, 2004; Munjiza et al., 2011; Munjiza et al., 2014). FDEM allows explicit geometric and mechanical realization of systems involving both continua and discontinua, which makes it superior to both FEM and DEM. Since its inception (Munjiza, 1992), FDEM has proven its computational efficiency and reliability, and has been extensively used in a wide range of endeavors in both industry and academia (Euser et al., 2019; Gao et al., 2018; Lei and Gao, 2018; Lei et al., 2019; Rougier et al., 2019). Additionally, benefitting from the recent implementation of a large-strain large-rotation formulation and grand scale parallelization in FDEM by the Los Alamos National Laboratory (Lei et al., 2014; Munjiza et al., 2014), the FDEM software package – HOSS (Hybrid Optimization Software Suite) (Knight et al., 2015; Munjiza et al., 2013) – offers a powerful tool to study the behavior of sheared granular fault system.

## 2.2. Model setup

Figure 1a illustrates the geometry of the FDEM model, which is based on the two-dimensional photoelastic laboratory experiment conducted by Geller et al. (2015). The model consists of 2,817 cylindrical particles confined between two identical deformable plates. The diameter of the particles is either 1.2 or 1.6 mm, and they are of similar amount and randomly placed between the plates. Each plate has dimensions of 570 mm × 250 mm in width and height, respectively. Two stiff bars are attached to the bottom end of the lower plate and to the top end of the upper plate on which the normal load $P$ and shear velocity $V$ are applied, respectively. At the interfaces between the plates and the particles, a series of half-circular shaped protuberances or "teeth" are added to the plates (Figure 1a). The teeth diameter and the separation between them are 1.6 mm and 0.8 mm, respectively. A number of "sensor" points are set on the centers of both the upper and lower teeth, close to the interface with the gouge, to track the motion of both plates with respect to stick-slip. To avoid edge effects, a portion comprising around 80 mm on both left and right sides of the gouge are not considered in the plate motion analysis, and thus the information collected from a total of 286 sensors (143 on each plate) is used in this work. This data collection reflects the dynamic evolution of a section of about 340 mm in length located in the middle area of the gouge (Figure 1a).

The granular fault gouge is consolidated at the beginning by moving the top and bottom stiff bars towards each other to ensure the particles are well contacted. After the consolidation, the gouge thickness and length are around 11.7 mm and 500 mm, respectively. Then the top stiff bar begins shearing, i.e., displacing horizontally towards the right-hand side, with a constant horizontal velocity *V*, while the normal load *P* on the bottom stiff bar is kept constant throughout the simulation. During the shearing stage the top stiff bar is allowed to move only in the *x* direction and the bottom stiff bar is allowed to move only in the *y* direction. The main simulation parameters are tabulated in Table 1, while a detailed illustration of the model geometry and parameter selection is presented in Text S2 of the Supplementary Material. The simulation results were compared and calibrated against the laboratory experiments in Gao et al. (2018), which shows good agreement in terms of seismic moment and thus demonstrates the capability and accuracy of FDEM for such simulation.

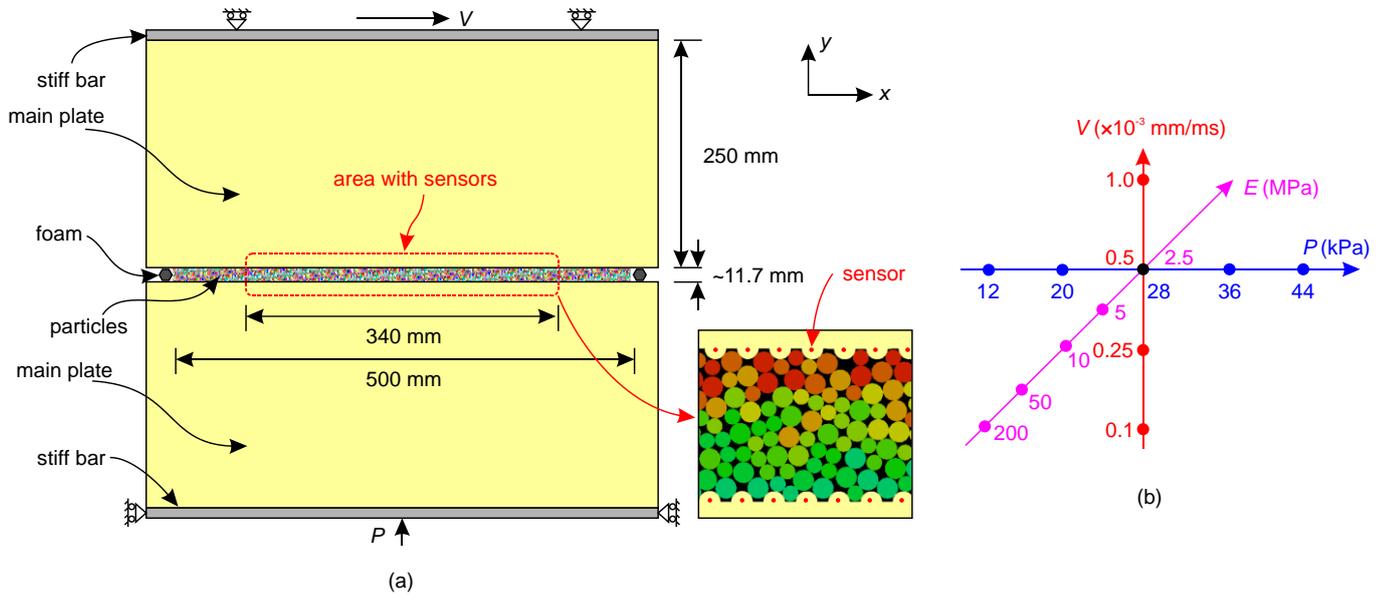

Figure 1. Model setup and domain of key simulation parameters. (a) FDEM model of the granular fault system and the sensor locations for plate motion monitoring. (b) Three-axis illustration of the three groups of models for examining the influence of normal load *P* (first group, blue), shear velocity *V* (second group, red) and plate Young's modulus *E* (third group, magenta) on plate motion; the common parameter combination of the three model groups is marked as black.

Table 1. Material and numerical simulation parameters.

| Property | Value | Property | Value |
| --- | --- | --- | --- |
| Particle diameter | 1.2 or 1.6 mm | Stiff bar density | 2,800 kg/m$^3$ |
| Particle density | 1,150 kg/m$^3$ | Stiff bar Young's modulus | 30 GPa |
| Particle Young's modulus | 0.4 GPa | Stiff bar Poisson's ratio | 0.33 |
| Particle Poisson's ratio | 0.4 | Foam density | 1,150 kg/m$^3$ |
| Particle-particle friction coefficient | 0.15 | Foam Young's modulus | 1 MPa |
| Number of particles | 2,817 | Foam Poisson's ratio | 0.4 |
| Main plate density | 1,150 kg/m$^3$ | Contact penalty | 4 GPa |
| Main plate Poisson's ratio | 0.49 | Time step | 1.0E-4 ms |
| Particle-plate friction coefficient | 0.15 | | |

In the following simulations, a shear velocity $V$ = 5.0E-4 mm/ms is selected first, and for the models with plate Young's modulus $E$ = 2.5 MPa, a series of normal loads $P$ ranging from 12 kPa to 44 kPa with increment of 8 kPa are used to investigate the influence of the normal load on plate motion. Then, to explore the effect of shear velocity, for the models with the same plate Young's modulus $E$ = 2.5 MPa and subjected to $P$ = 28 kPa, another three shear velocities $V$ = 1.0E-4, 2.5E-4 and 1.0E-3 mm/ms are employed. Finally, for the models subject to $P$ = 28 kPa and $V$ = 5.0E-4 mm/ms, another four plate Young's modulus $E$ = 5, 10, 50 and 200 MPa are employed to further examine the influence of plate stiffness on simulation results. Here, we only focus on the effect of normal load, shear velocity and plate stiffness on plate motion, with all other parameters being kept the same for all models. All combinations of normal load, shear velocity and plate stiffness used here are within the domain that can guarantee the generation of typical repetitive stick-slip events. The parameter combination of the three groups of models are summarized in Figure 1b.

The simulations use a time step of 1.0E-4 ms, and each model is run for roughly 3.0E+8 time steps with a total shearing time of approximately 30,000 ms. Each model reaches steady state after the first 3,000 ms approximately. Because of this, in this work the data for the analysis was collected after 5,000 ms of simulation time. The shear and normal forces between the particles and the upper and lower plates, as well as the $x$ velocity and displacement of the 143 pairs of sensors, are recorded every 1 ms. This time step interval for output recording is chosen carefully through a series of comparisons by considering the resolution of output as well as the computational cost. The particle-plate shear and normal forces are calculated by first resolving the normal and tangential contact forces between each particle-plate contact pair into $x$ and $y$ directions and then integrating them respectively along the particle-plate interfaces. The ratio of the shear to normal force is then calculated as the macroscopic friction coefficient between the plates and granular fault gouge.

## 3. Simulation results

### 3.1. General characteristics of plate motion

The $x$ velocities of the 143 sensor points on each of the upper and lower plate are averaged at each output step. The time series of the averages for the model with the common parameter combination indicated by the black dot in Figure 1b (i.e., $P$ = 28 kPa, $V$ = 5.0E-4 mm/ms, and $E$ = 2.5 MPa) are presented in Figure 2. The average $x$ velocity of the upper plate sensors exhibits primarily positive values, whereas the lower plate sensors generally exhibit the opposite. The upper and lower $x$ velocities are anti-correlated. To facilitate comparison between the plate motion and the stick-slips in the granular fault system, we also plot in Figure 2 the change of gouge-plate macroscopic friction coefficient with respect to time. During stick phases, the macroscopic friction coefficient increases in an approximately linear manner. At the end of a stick phase, a rapid drop of macroscopic friction coefficient, which marks a slip event, can be observed. The temporal variations of both the upper and lower plate average *x* velocities match well with the gouge-plate macroscopic friction coefficient. When slip occurs, because of the partial contact loss between plates and particles, the bottom portion of the upper plate lurches to the right and the top portion of the lower plate resets towards its original position to the left. As a result, the upper and lower gouge-plate interfaces have a simultaneous sudden increase of *x* velocity magnitude but in opposite directions. Particularly, large *x* velocity magnitudes are generally associated with large macroscopic friction coefficient drops. Additionally, the average *x* velocities of the upper and lower plate sensors with respect to time are nearly symmetric, manifested by a spike at the bottom of the upper plate towards the right, attended by a spike of similar magnitude at the top of the lower plate toward the left. Whereas during the stick phases, a gradual increase of macroscopic friction coefficient is regularly accompanied with a constant average *x* velocity around 2.5E-4 mm/ms (half of the shear velocity *V*) in the direction of shear for both upper and lower plates (see inset of Figure 2). All other models present the similar general characteristic of *x* velocity in both stick and slip phases.

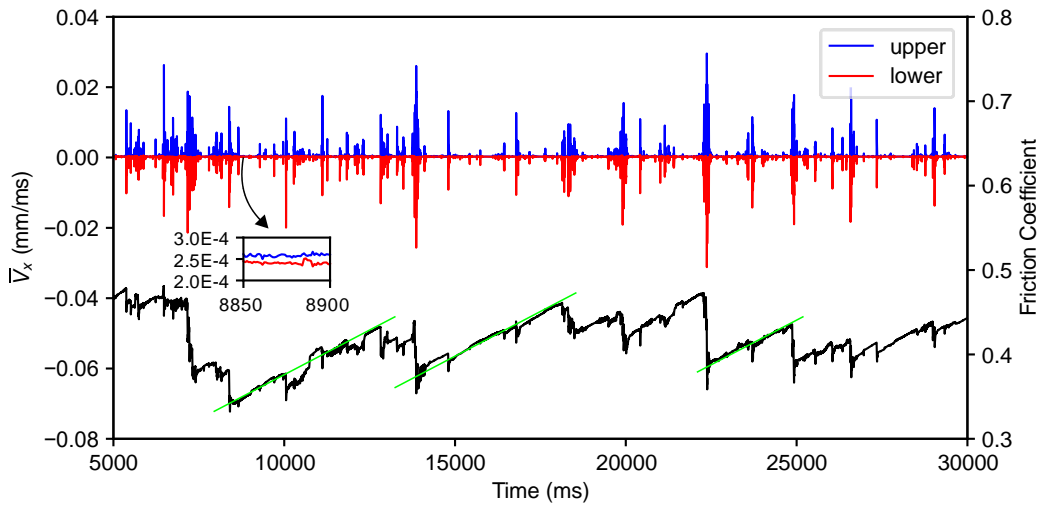

Figure 2. Comparison between the average *x* velocity at places adjacent to the plate-gouge interfaces and the plate-gouge macroscopic friction coefficient. Upper panel (left *y* axis): time series of average *x* velocity of the 143 sensors on the upper (blue) and lower (red) plate, respectively; the inset illustrates the average velocity difference between the upper and lower plates during the stick phases. Lower panel (right *y* axis): time series of macroscopic friction coefficient; the green lines indicate the approximately constant slope of the friction coefficient.

To further interpret plate motion, we examine the average *x* displacements ($\bar{D}_x$) of the 143 upper and 143 lower plate sensors in each output step, and their time series are presented in Figure 3a. The average *x* displacements of upper plate sensors move from 0 to 13.5 mm approximately uniformly in time with an average velocity $\bar{V} \approx 13.5/27000 = 5.0\text{E-4}$ mm/ms, which is equal to the shear velocity *V* on the upper plate. The average *x* displacements of the lower plate sensors stay around $x = 0$, i.e., the lower plate does not on average move. In stick phases, the upper and lower plates are locked and moving together with a velocity around $V/2$.

At slip time, the top of the lower plate retreats to $x \approx 0$ and the bottom of the upper plate lurchers to a new position atop the lower plate. The upper and lower plates are then locked anew and repeat the stick-slip cycles, as is demonstrated in the cartoon in Figure 3b. In the following two sections, we give a detailed analysis of the plate motion in terms of *x* velocities at the sensors in slip and stick phases, respectively, and investigate how they are influenced by the normal load, shear velocity and plate stiffness.

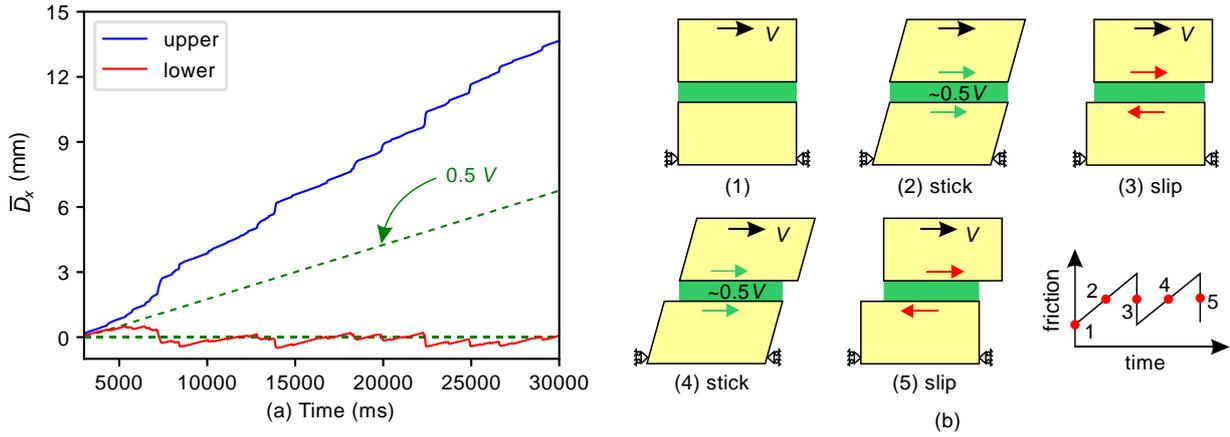

Figure 3. (a) Time series of average *x* displacement of the 143 sensors on the upper (blue) and lower (red) plate, respectively; the two green auxiliary dotted lines indicate the displacements having zero and $V/2$ velocities, respectively. (b) Cartoon illustration of plate motion during stick and slip phases.

### 3.2. Plate motion at slips

Because of the negative correlation between the *x* velocities of upper and lower plate sensors during slip phases and by considering the results of both, we use a "couple strength" of each upper and lower sensor pair to explore the plate motion along the gouge-plate interfaces. The *x* velocity couple strength for each sensor pair in a specific output time step is defined as

$$V^C_{x_i} = \frac{V^U_{x_i} - V^L_{x_i}}{2}, \qquad (2)$$

where the superscripts "*C*", "*U*" and "*L*" denote "couple", "upper" and "lower", respectively, and $i = 1,2,\ldots 143$ is the sensor number counted from left to right along the gouge for both the upper and lower plates. Since here we focus on plate motions at slip, only $V^C_x \geq$ 1.0E-3 mm/ms are considered. The complementary cumulative distribution functions (CCDFs) of $V^C_x$ for the three groups of models are presented in Figure 4. The CCDF gives the probability of an *x* velocity couple strength larger than or equal to a certain magnitude, and thus provides a useful tool for comparing the influence of different parameters on the frequency and magnitude of slip events, especially on the large ones. As can be seen from Figure 4, the largest *x* velocities of the three groups of models are approximately located around 0.1 mm/ms, and could be as large as 2-3 orders of the shear velocity. As the normal load increases, larger $V_x^C$ are generated, which indicates that the system is more responsive at slips when subjected to higher normal loads (Figure 4a). The shearing velocity seems to have no significant influence on plate motion along the gouge-plate interfaces at slips, as is manifested by the nearly overlapping CCDFs in Figure 4b. However, stiffer plates suppress local vibrations and result in a relatively "quiet" system with smaller *x* velocity magnitudes at slips (Figure 4c).

The *x* velocity couple strength correlates well with the drop of macroscopic friction coefficient. To compare the two, we first average the 143 *x* velocity couple strengths in each output step, and then for the time intervals at slips we calculate the drop of friction coefficient and also average the calculated mean couple strengths between adjacent output time steps. The results are plotted in Figure 5, which generally demonstrates that slips with larger *x* velocity couple strengths are usually associated with larger drops of macroscopic friction coefficient. Moreover, for the models subjected to smaller normal loads, a similar *x* velocity couple strength can cause larger drops of macroscopic friction coefficient (Figure 5a). This is mainly because of the relatively smaller contact area between gouge and plates when a smaller external normal load is employed. As a result, when slip occurs, even a small contact ratio drop can cause large friction loss. Likewise, stiffer plates yield less gouge-plate contact area when compared with the cases with softer plates; therefore, larger friction coefficient drop could occur for the models with stiffer plates (Figure 5c). Similar as before, the shear velocity *V* has no obvious effect on the change of correlation between macroscopic friction coefficient drop and *x* velocity couple strength (Figure 5b).

Both Figure 4 and Figure 5 generally reveal that the maximum values of $V_x^C$ increase with the increasing normal load and shear velocity (although relatively small), as well as with decreasing plate Young's modulus. This is mainly because when the model is subjected to larger normal loads, higher amounts of strain energy are stored in the system during the stick phases, and when slip occurs this strain energy is released in the form of larger vibrations. For the model sheared with larger velocity, because more work is done on the system per unit time, the system is thus more energetic and more vibrations are observed during the shear phases. However, as mentioned earlier, stiffer plate constrains the vibration and results in a system with smaller velocity couple strengths.

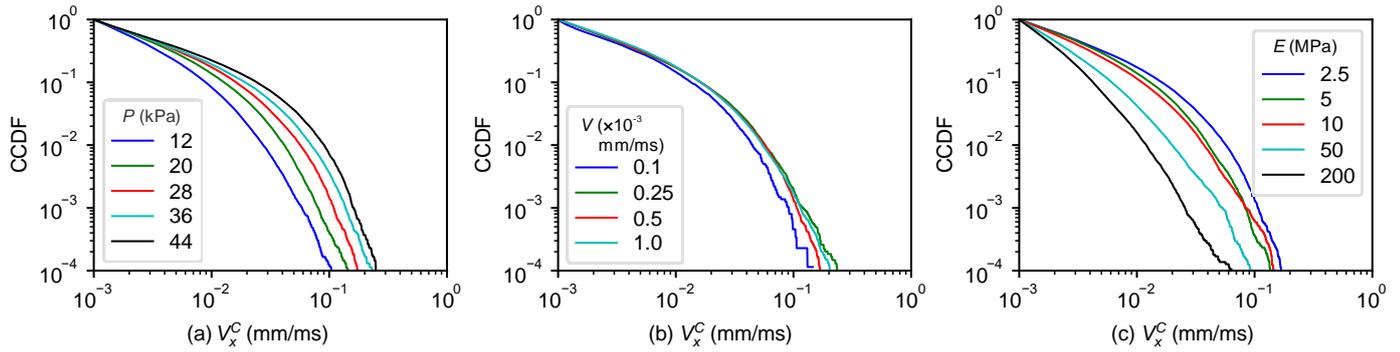

Figure 4. Complementary cumulative distribution functions (CCDFs) of the *x* velocity couple strength during slips for the (a) first, (b) second and (c) third group of models demonstrated in Figure 1b.

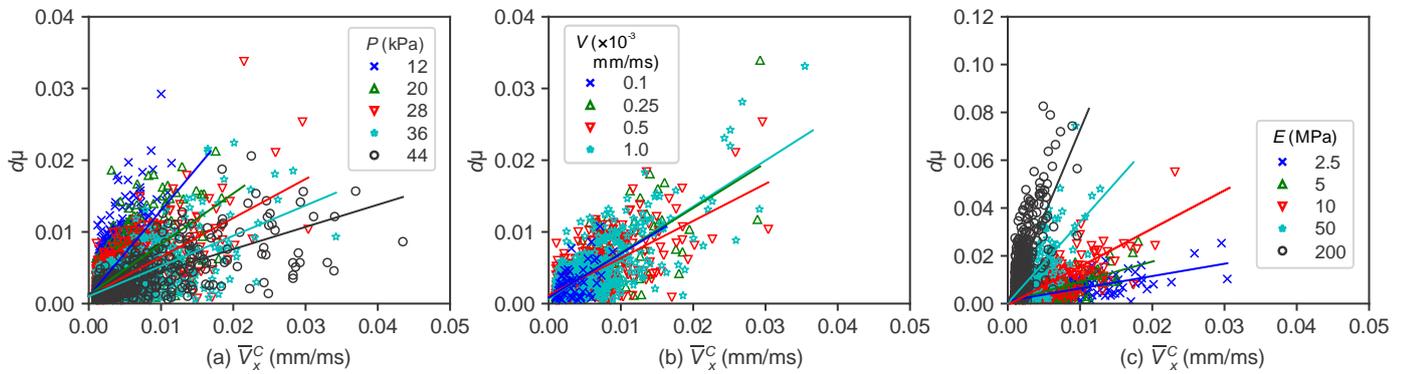

Figure 5. Relationship between the drop of macroscopic friction coefficient and the corresponding *x* velocity couple strength for the (a) first, (b) second and (c) third group of models demonstrated in Figure 1b. The lines are linear fits between the friction coefficient drop and the *x* velocity couple strength to facilitate comparison.

### 3.3. Plate motion during stick phases and interpretation of gouge shear strength

To examine plate motion during stick phases, we first plot in Figure 6a the probability density functions (PDFs) of all the output *x* velocities of the 143 sensor points on each of the upper and lower plates for the model with the common parameter combinations indicated by the black dots shown in Figure 1b. The markers with the same color as the line represent the corresponding maximum and minimum *x* velocities. As can be seen from Figure 6a, the *x* velocities have both positive and negative values, and positive velocities with large magnitude are mainly seen on the upper plate, while negative velocities with large magnitude mostly occur on the lower plate. This is consistent with the signs of average *x* velocities shown in Figure 2. The *x* velocities of the sensors on both plates are highly concentrated around 2.5E-4 mm/ms (i.e., $V/2$, see inset of Figure 6a), and the modes of PDFs shown in the inset reflect plate motions during the stick phases. A quick comparison reveals that during the stick phases, the upper plate sensors *x* velocities are slightly greater than $V/2$ and the lower plate sensors *x* velocities are slightly less than $V/2$, which is in response to the average *x* velocity difference between the two plates shown in the inset of Figure 2. This *x* velocity difference is an evidence that the gouge is experiencing a growing shear strain necessary to support the growing shear stress in the system. Specifically, during stick phases the shear stress in the system is growing steadily (as indicated in Figure 2), and the gouge must support the growing shear stress; it does so by developing a growing shear strain, i.e., the upper plate in places adjacent to the gouge moves slightly faster than that of the lower plate. The plate motions of other models, shown in Figures S3-S5 in the Supplementary Material, manifest similar characteristics.

The velocity difference between the upper and lower plate sensors, i.e.,

$$dV_x^S = V_x^{U'} - V_x^{L'}, \qquad (3)$$

indicates the rate of shear strain growth across the gouge, where the superscript "$S$" denotes "stick", and $V_x^{U'}$ and $V_x^{L'}$ represent the upper and lower plate $x$ velocity during stick phases (i.e., the modes of PDFs of $x$ velocities), respectively. We calculate $dV_x^S$ for the three groups of models and the results are presented in Figure 6b-d: $dV_x^S$ decreases with the increasing normal load, and increases with the increasing shear velocity and plate Young's modulus. The shear strain rate can be further calculated by

$$\dot{\varepsilon} = \frac{d\varepsilon}{dt} = \frac{dV_x^S}{\bar{H}}, \qquad (4)$$

where $\bar{H}$ is the average gouge thickness in each model, and here $\bar{H} \approx 11.7$ mm. As can be interpreted from Figure 6b-d, large normal loads suppress the shear strain rate, which reveals that during stick phases the gouge is more compact and the upper and lower plates are moving in a more synchronized fashion (i.e., smaller relative velocity difference) when the system is compressed by larger normal loads. However, a faster shear velocity increases the relative velocity difference between the upper and lower plates and thus yields a larger shear strain rate. When the shear plates are stiffer, deformation is limited inside the plates and thus, the bottom part of the upper plate moves in an almost similar manner as the top part in which the constant shear velocity is enforced; since the bottom part of the lower plate is fixed in $x$ direction, more sliding and particle rolling will occur during the shear and thus larger shear strain rate is generated. Particularly, the plate stiffness has more notable influence on the shear strain rate than the normal load and shear velocity, as is demonstrated by the larger $dV_x^S$ in Figure 6d when compared with Figure 6b & c.

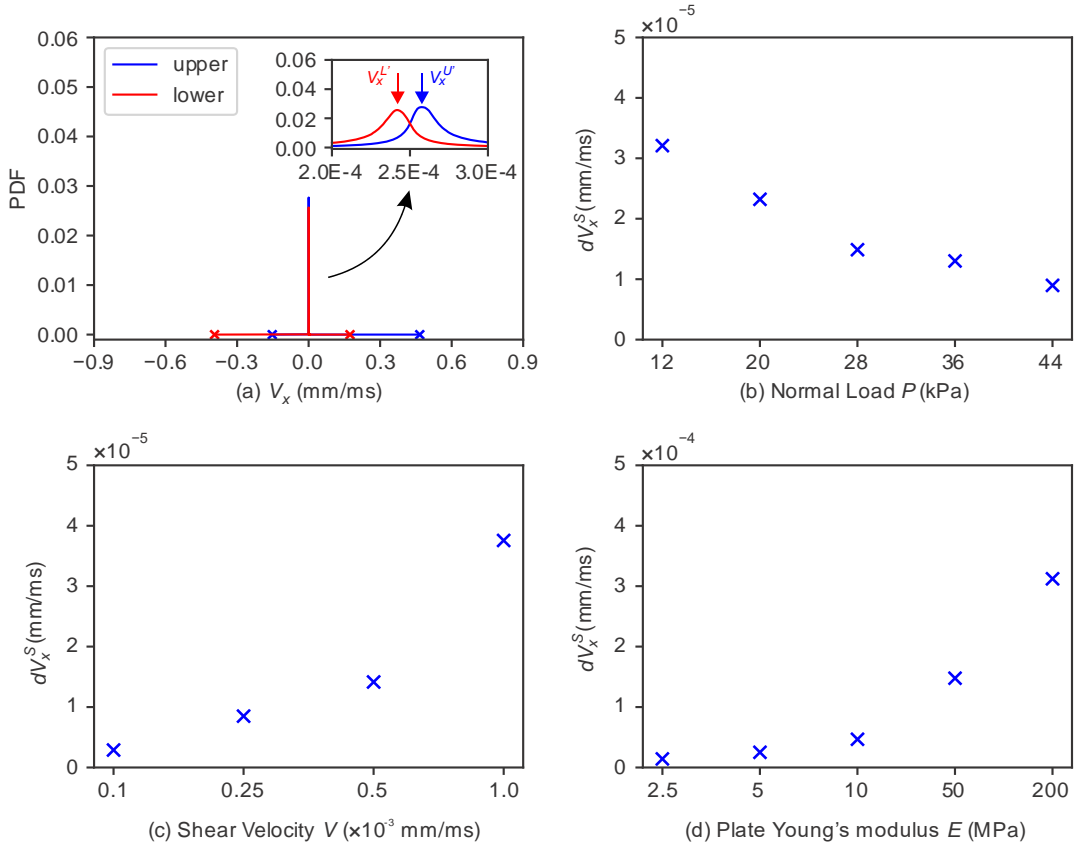

Figure 6. Analyses of $x$ velocity of the 143 sensor pairs during stick phases. (a) Probability density functions of all output $x$ velocities of the 143 sensor pairs on the upper and lower plate, respectively, for the model with the common parameter combination of normal load, shear velocity and plate Young's modulus shown in Figure 1b; the markers with the same color as the line represent the corresponding maximum and minimum values. The velocity difference between the upper and lower plates during the stick phases for the (a) first, (b) second and (c) third group of models demonstrated in Figure 1b.

Additionally, as mentioned earlier and indicated by the green lines in Figure 2, the macroscopic friction coefficient between the gouge and plates also advances in an approximately uniform rate in time. This gives an indication of shear stress rate applied on the gouge, as we have

$$\dot{\tau} = \frac{d\tau}{dt} = P\frac{d\mu}{dt}, \tag{5}$$

where $\tau$ is the shear stress between gouge and plates and $\mu$ is the macroscopic friction coefficient. Thus, we can obtain the effective shear modulus of the gouge for each model in an approximation manner using

$$G = \frac{d\tau}{d\varepsilon} = \frac{\dot{\tau}}{\dot{\varepsilon}}. \tag{6}$$

The gouge effective shear moduli of the three groups of models are calculated and presented in Figure 7. These effective shear moduli are generally within the range between 0.2 and 0.9 MPa. The increasing normal load enhances the effective shear modulus of the granular gouge (Figure 7a). While the gouge effective shear modulus reduces when subjected to larger shear velocity (Figure 7b). A stiffer shear plate could increase the gouge effective shear modulus (Figure 7c). All these may be attributed to the influence of normal load, shear velocity and plate Young's modulus on the stress chain distributions inside the granular gouge (Gao et al., 2019). Specifically, larger normal load intensifies the stress chains formed by adjacent particle interactions and thus increases the gouge effective shear modulus. Likewise, a stiffer shearing plate can more effectively transfer external normal load to the gouge and strengthen particle contacts. However, a faster shear velocity may weaken the stress chain network and reduce the gouge effective shear modulus. Furthermore, the effective shear modulus increases approximately linearly with the increasing normal load, e.g., a nearly four times increase of normal load (from 12 kPa to 44 kPa) results in a roughly four times rise of gouge effective shear modulus (Figure 7a). While compared with the normal load, the effect of shear velocity and plate Young's modulus on gouge effective shear modulus is less significant. For instance, ten-times increase of shear velocity only results in a less than 10% reduction of gouge shear modulus, and similarly, the gouge effective shear modulus only witnesses 10% increases when the plate stiffness is increased by 80 times.

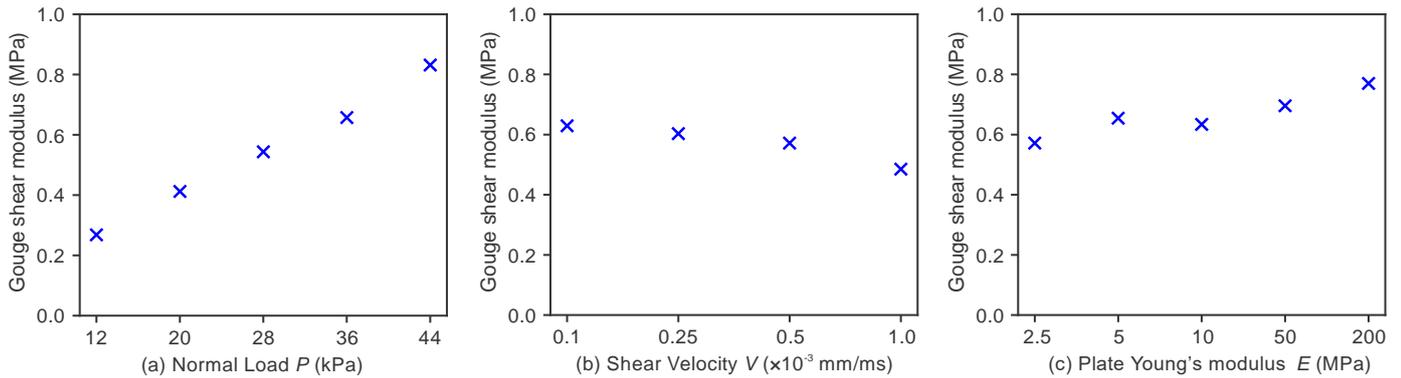

Figure 7. Gouge effective shear modulus of the (a) first, (b) second and (c) third group of models demonstrated in Figure 1b.

## 4. Discussion

The two main characteristics derived here, i.e., the *x* velocity couple strength which represents the plate motion at slip times and the gouge effective shear modulus interpreted from the motion of the plates during stick phases, may be critical for revealing the complex mechanism of earthquake. As discussed above, these two characteristics are influenced by the normal loads, shear velocities and plate Young's modulus. Here, a potential scaling for these two characteristics with respect to the influencing factors is discussed. First, for the first group of models, after dividing the *x* velocity couple strengths at slips by the corresponding normal load, the resulting CCDFs are close to each other (Figure 8a). This reveals that the plate motion at slips is proportional to the normal load acting on the system. While for the second group of models, the four shear velocities do not have very distinct influence on the plate motion during slips (see Figure 4b). Regarding the influence of different plate Young's modulus, we found that the CCDFs of the *x* velocity couple strengths of the third group of models multiplied by $\sqrt{E}$ seems reasonably close (Figure 8b). This indicates that the plate motion at slips may be inversely proportional to the square root of plate Young's modulus.

Similar to the effect of normal load on plate motion at slips, the gouge effective shear modulus also scales well with the normal load (Figure 8c), i.e., gouge shear modulus increases proportionally with the normal load. Therefore, we suspect the geometrical structure of stress chains that carry forces are similar for all models with different normal loads and expect the spectrum of stresses to have amplitudes controlled by the normal load. This needs further detailed systematic investigation. However, although the shear velocity and plate stiffness do influence the gouge effective shear modulus, e.g., larger shear velocity could weaken the gouge effective shear modulus and a stiffer plate may enhance the gouge effective shear modulus, their influence is minor, especially when compared to the influence by the normal load. Through the above analyses, we speculate that the normal load determines the structure of elastic elements in the gouge that give the system its frictional properties. In other words, the normal load determines the nature and strength of the stress chains. Whereas the shear velocity and Young's modulus of plates are the driving system which determines the rate at which the elastic elements in the gouge have their strength tested.

In the current analysis, only the influence of normal load, shear velocity and plate stiffness on plate motion is discussed. However, other parameters such as gouge thickness, particle-particle and particle-plate friction coefficient, particle size distribution and the presence of fluids may also play a significant role on the behavior of elastic elements in the gouge that give the system its frictional properties. Further work is necessary to provide a thorough exploration of this.

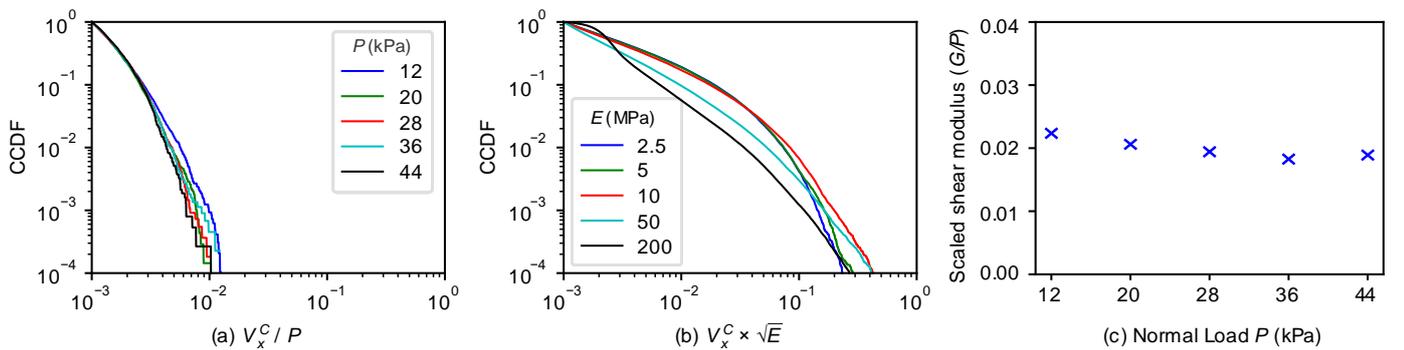

Figure 8. Scaling of the results related to plate motion in stick and slip phases: scaling of CCDFs of the *x* velocity couple strength during slips for the (a) first, (b) third group of models; (c) scaling of gouge shear modulus for the first group of models.

# 5. Conclusions

We have used the FDEM to explicitly simulate the stick-slip induced near gouge plate motion in sheared granular fault systems. In the FDEM model, the plates and particles are represented by discrete elements to track their motion and interaction with neighboring objects, and each discrete element is further discretized into finite elements to capture its deformation during shear loads. Three groups of models for a total of 12 simulations have been conducted to respectively explore the influence of normal load, shear velocity and plate stiffness on the plate motion in terms of $x$ velocity and $x$ displacement in places adjacent to the gouge during the stick and slip phases. The gouge effective shear modulus is also interpreted based on the plate motion, and a potential scaling of plate motion at slips and the gouge effective shear modulus with respect to normal load, shear velocity and plate stiffness is discussed.

The simulations show that for all the models the $x$ velocities of the 143 sensor points on each of the upper and lower plates have both positive and negative values. The positive $x$ velocities with large magnitude are mainly observed on the upper plate, while the negative $x$ velocities with large magnitude mostly occur on the lower plate. The average $x$ velocity analysis demonstrates that at slips the upper and lower gouge-plate interfaces have a simultaneous spike of $x$ velocity with similar magnitude but in opposite directions. These indicate that when slip occurs the bottom of the upper plate lurches to the right and the top of the lower plate resets towards the left. During the stick phases, both upper and lower plate sensors move in a constant average $x$ velocity around half of the shear velocity in the direction of shear. The $x$ displacement analysis shows that on average the upper plate sensors move almost uniformly in time with an average $x$ velocity equal to the shear velocity on the upper plate, whereas the lower plate does not on average move.

The $x$ velocity couple strength of each sensor pair on the upper and lower plates is used to analyze the plate motion during slip. The results show that the maximum values of the $x$ velocity couple strength increase with the increasing normal load and shear velocity as well as the decreasing plate Young's modulus, and the maximum $x$ velocities could be as large as 2-3 orders of the shear velocity. As the normal load increases, larger $x$ velocity couple strengths are generated. The shearing velocity seems to have no significant influence on plate motion at slips, while stiffer plates result in a relatively "quiet" system with smaller $x$ velocity magnitudes. The $x$ velocity couple strength correlates well with the drop of macroscopic friction coefficient, and large $x$ velocity magnitudes are generally associated with large macroscopic friction coefficient drops. For the models subjected to smaller normal loads, a similar $x$ velocity couple strength can cause larger drops of macroscopic friction coefficient. Large friction drops can also be more frequently observed at slips when using stiffer plates. However, the shear velocity seems to have no distinct effect on the correlation between macroscopic friction coefficient drop and $x$ velocity couple strength.

During the stick phases, the upper and lower plate sensors $x$ velocities are respectively slightly greater and less than the half of the shear velocity. This small $x$ velocity difference between the upper and lower plates adjacent to the gouge during stick phases is an evidence that the gouge is experiencing a growing shear strain necessary to support the growing shear stress in the system. We have calculated both the shear strain and shear stress rates for all the models and thus obtained the gouge effective shear modulus. The results show that both the increasing normal load and plate stiffness could enhance the effective shear modulus of the gouge, while the gouge effective shear modulus reduces when subjected to larger shear velocity. However, compared with the normal load, the effect of shear velocity and plate Young's modulus on gouge effective shear modulus is less significant.

The scaling analysis shows that the *x* velocity at slips is proportional to the normal load acting on the system. The shear velocity does not show very distinct influence on the plate motion at slips. The *x* velocity at slips may be inversely proportional to the square root of plate Young's modulus. Additionally, the gouge effective shear modulus increases proportionally with the normal load, while the influence of shear velocity and plate stiffness on gouge effective shear modulus is minor when compared with the normal load. We suspect that the normal load determines the nature and strength of the stress chains that mainly control the frictional properties of the gouge. Whereas the shear velocity and Young's modulus of plates only determines the rate at which the elastic elements in the gouge have their strength tested. The simulations disclose the influence of normal load, shear velocity and fault block stiffness on the stick-slip induced near gouge vibration, and may be helpful for understanding the complex behavior of earthquake source physics and dynamics.

## Acknowledgements

Institutional Support [LDRD] at the Los Alamos National Laboratory (50%) and the Office of Science, Basic Energy Science (50%) supported this work. Technical support and computational resources from the Los Alamos National Laboratory Institutional Computing Program are highly appreciated.